\documentclass[aps,pra,showpacs,floatfix]{revtex4}
\usepackage{graphicx}
\usepackage{times}
\usepackage{nicefrac}
\usepackage{amsmath}
\usepackage{amsfonts}
\usepackage{amssymb}
\usepackage{amsthm}
\usepackage{epsf}
\usepackage{bm}
\usepackage{bbm}
\usepackage{color}

\usepackage{dcolumn}
\newcolumntype{.}{D{x}{}{-1}}

\newcommand{\be}{\begin{eqnarray}}
\newcommand{\ee}{\end{eqnarray}}
\newcommand{\la}{\langle}
\newcommand{\ra}{\rangle}

\newcommand{\veps}{\varepsilon}

\newcommand{\pr}{\prime}

\newcommand{\balpha}{\bm{\alpha}}

\newcommand{\beps}{\bm{\epsilon}}

\newcommand{\bfk}{{\bf k}}

\newcommand{\bfr}{{\bf r}}

\begin{document}
%
\title{Many-electron effects on the x-ray Rayleigh scattering by highly charged He-like ions}
\author{A. V. Volotka,$^{1,2}$ V. A. Yerokhin,$^{3}$ A. Surzhykov,$^{1}$ Th. St\"ohlker,$^{1,4,5}$ and S. Fritzsche$^{1,5}$}

\affiliation{
$^1$ Helmholtz-Institut Jena, D-07743 Jena, Germany\\
$^2$ Department of Physics, St. Petersburg State University, 198504 St. Petersburg, Russia\\
$^3$ Center for Advanced Studies, Peter the Great St. Petersburg Polytechnic University, 195251 St. Petersburg, Russia\\
$^4$ GSI Helmholtzzentrum f\"ur Schwerionenforschung, D-64291 Darmstadt, Germany\\
$^5$ Theoretisch-Physikalisches Institut, Friedrich-Schiller-Universit\"at Jena, D-07743 Jena, Germany\\
}

\begin{abstract}
The Rayleigh scattering of x-rays by many-electron highly charged ions is studied
theoretically. The many-electron perturbation theory, based on a rigorous quantum
electrodynamics approach, is developed and implemented for the case of the elastic
scattering of (high-energetic) photons by helium-like ion. Using this elaborate
approach, we here investigate the many-electron effects beyond the independent-particle 
approximation (IPA) as conventionally employed for describing the Rayleigh 
scattering. The total and angle-differential cross sections are evaluated for the 
x-ray scattering by helium-like Ni$^{26+}$, Xe$^{52+}$, and Au$^{77+}$ ions
in their ground state. The obtained results show that, for high-energetic photons,
the effects beyond the IPA do not exceed 2\% for the scattering by a
closed $K$-shell.
\end{abstract}

\maketitle

\section{Introduction}
\label{sec:1}

The elastic scattering of a light by bound electrons is commonly known as
Rayleigh scattering. It has been found a powerful and versatile tool for
investigating the structure and dynamics of bound electrons as well
as for probing the atomic environment. Apart from the fundamental interest,
a quantitative understanding of the Rayleigh scattering is needed in 
various fields, including the study of the solid-state, complex molecules
or even nano-objects \cite{sfeir:2004:1540,kampel:2012:090401,kulik:2012:113403,
wu:2015:303}, astrophysics \cite{maeda:2012:54,the:2014:141} as well as in
medical diagnostics \cite{zhang:2012:6408}.

Experimentally, recent progress in exploring the Rayleigh scattering of hard x rays by
atomic or solid-state targets has been achieved by a major improvement of the detection
techniques as well as the quality of light sources. Indeed, novel solid-state photon
detectors \cite{fritzsche:2005:S707,weber:2010:C07010} together with advances in the
available synchrotron sources \cite{bilderback:2005:S773} have paved the way towards new
generations of experiments. For example, a measurement of the linear polarization of the
Rayleigh scattered light has been recently performed with the help of the segmented
solid-state detectors at the PETRA III synchrotron at DESY \cite{blumenhagen:tbp}. The
advances in the Rayleigh scattering experiments nowadays also demand further and accurate
predictions at the side of theory.

The theoretical investigations on the elastic scattering of photon
by bound electrons dates back to the mid-1930s \cite{franz:1936:314}. While initially
quite simple approximations were applied, based on the atomic form factors, the rigorous 
quantum electrodynamical (QED) approach has later been developed by using the
relativistic second-order $S$-matrix amplitude; we refer the reader to Ref.~\cite{kissel_pratt}
for a comprehensive historical overview. This latter approach has now become the standard
for treating the Rayleigh scattering. These developments were triggered specially
by the pioneering work of Brown, Peierls, and Woodward \cite{brown:1954:51}, who
developed the method for the calculation of the second-order transition amplitudes. 
Within this QED approach, quite a number of calculations were carried out 
for different atoms and photon energies 
\cite{brenner:1954:59,brown:1956:387,brown:1957:89,johnson:1976:692,kissel:1980:1970,
kane:1986:75,roy:1986:1178}; we refer the reader to Ref.~\cite{roy:1999:3} for a
comprehensive review on this approach. In recent years
\cite{cortescu:2011:045204,safari:2012:043405,surzhykov:2013:062515,surzhykov:2015:144015,safari:2015:271} 
the angular and polarization correlation between the incident and outgoing photons have also
been investigated. Unfortunately, however, the formalism of the second-order $S$-matrix 
theory does not enable one to investigate systematically the many-electron effects
on the Rayleigh scattering. This is caused by the electron-electron
interaction that is treated only approximately in this formalism by means of a
central screening potential, and which is known as screening potential or
independent-particle approximation (IPA). Until the present, many-electron effects
beyond the IPA were only investigated for the helium atom in Ref.~\cite{lin:1975:1946}.
It was found that the interelectronic-interaction corrections beyond the IPA significantly
modify the Rayleigh cross section for the low-energy photons and disappear for higher
energies. The main aim of the present study is to investigate the many-electron effects
for highly charged ions.

In this paper, we present a rigorous QED treatment of the Rayleigh scattering of light
by highly charged helium-like ions. The QED perturbation expansion with regard
to the interelectronic interaction is applied up to the first-order for the 
Rayleigh scattering as an important light-matter interaction process at medium and
high photon energies. In particular, formulas have been derived for the zeroth- and first-order
interelectronic-interaction corrections to the scattering amplitude. This framework enables
one to systematically investigate the many-electron effects beyond the IPA. In Sec.~\ref{sec:2}, 
the details of this formalism are presented, while the computational techniques
and methods are discussed in Sec.~\ref{sec:3}. The total and angle-differential
cross sections are evaluated for the scattering of x rays on the ground state of 
helium-like Ni$^{26+}$, Xe$^{52+}$, and Au$^{77+}$ ions. We shall here consider 
especially two experimental scenarios in which the incoming light is either unpolarized 
or completely linearly polarized. In Sec.~\ref{sec:4}, the obtained numerical results are 
presented and discussed. Attention is paid to the comparison between the IPA and the 
many-electron description. For the Rayleigh scattering of x rays by helium-like ions in their
ground state, we find that the many-electron effects beyond the IPA treatment
do not exceed 2~\%. In Sec.~\ref{sec:5}, we briefly summarize our results and give a
brief outlook.

Relativistic units ($\hbar = 1,\,c = 1,\,m = 1$) and the Heaviside charge unit
[$\alpha = e^2/(4\pi)$, $e<0$] are used throughout the paper.

\section{Theoretical background}
\label{sec:2}

The process of elastic photon scattering is characterized by the energy conservation
of the incident and outgoing photons in the center-of-mass frame of the overall
scattering system. This means that no energy transfer is possible between the 
photon and the target with its internal degree of freedom. However, since the energy
of the scattered photon is typically much smaller than the atom's rest-mass energy,
we therefore consider in the following the scattering of a photon in the rest frame
of the atom, so that the incoming and outgoing photon energies simply remain the same.
For the theoretical description  of (quite) heavy atoms, moreover, it is naturally to
utilize the Furry picture, in which the (infinitely heavy) nucleus is taken as the
source of the classical Coulomb field and where the interaction of electrons with this
field is then treated exactly by just solving the Dirac equation in the nuclear Coulomb
potential.

According to the basic principles of QED \cite{berestetsky}, the differential cross
section for the scattering of a photon by an atom is given by
\be
d\sigma(k_f,\beps_f,A';k_i,\beps_i,A) = (2\pi)^4 |\tau_{\gamma_f,A';\gamma_i,A}|^2
  \delta(E_{A'} + k_f^0 - E_{A} - k_i^0)\: d\bfk_f\,,
\ee
where the initial and final state of the photon are characterized by the four-momentum
$k^\mu_i$ and $k^\mu_f$ and polarization vectors $\beps_i$ and $\beps_f$, respectively.
Here, the zero and spatial components of the four-vector define the photon frequency
$k^0 \equiv \omega$ and the photon wave vector $\bfk$. Moreover, the total energy of
the bound electrons are $E_{A}$ and $E_{A'}$ for the initial and final state 
of the atom. The shorthand notations $A$ and $A'$ stand for a unique
specification of the bound-electron states $A = \alpha_A J_A M_A$ and
$A' = \alpha_{A'} J_{A'} M_{A'}$, where $J_A$ and $J_{A'}$ are the total angular momenta,
$M_A$ and $M_{A'}$ their corresponding projections, and where $\alpha_A$ and 
$\alpha_{A'}$ denote all additional quantum numbers that are needed for a unique 
specification of the states. The energy conservation $E_{A} = E_{A'}$ clearly shows 
that no energy transfer occurs to the atom and that the moduli of the wave vectors
are the same for the incoming and outgoing photons, $k_i^0 = k_f^0$. Thus, 
the angle-differential cross section for the elastic scattering in a solid angle $d\Omega_f$ 
takes the form
\be \label{eq:2.0}
  d\sigma(\bfk_f;\bfk_i,\beps_i) =
  \frac{(2\pi)^4 (k_i^0)^2}{2J_A+1}\sum_{M_A,\,M_{A'}} \sum_{\beps_f}\:
  |\tau_{\gamma_f,A';\gamma_i,A}|^2 \; d\Omega_f\,.
\ee
The scattering amplitude $\tau_{\gamma_f,A';\gamma_i,A}$ can be related to the
scattering $S$-matrix element by following expression
\be \label{eq:2.1}
S^{\rm scat}_{\gamma_f,A';\gamma_i,A} \:=\: \la k_f,\beps_f,A'|(\hat{S}-\hat{I})|k_i,\beps_i,A\ra \:=\:
  2\pi i\;\tau_{\gamma_f,A';\gamma_i,A}\,\delta(k_f^0 - k_i^0)\,,
\ee
and where $\hat{S}$ and $\hat{I}$ denote the scattering and identity operators. 
The scattering $S$-matrix element $S^{\rm scat}_{\gamma_f,A';\gamma_i,A}$ contains two types of 
processes: the scattering by the Coulomb potential of the nucleus as well as the scattering
by the bound electrons.
The first type corresponds to the Delbr\"uck scattering amplitude, while the second one
is usually defined as the Rayleigh scattering by the bound electrons. Here, we restrict
ourselves to the Rayleigh scattering only. In order to evaluate the corresponding
Rayleigh $S$-matrix element, which we denote as $S^{\rm R}_{\gamma_f,A';\gamma_i,A}$,
one has to employ the bound-electron QED perturbation theory. For this purpose, we here
utilize the (so-called) two-time Green-function method as developed in 
Refs.~\cite{shabaev:1990:43,shabaev:1990:83,shabaev:2002:119}, where the perturbation theory 
is formulated for the two-time Green functions. The Rayleigh $S$-matrix element can be 
generally related to the two-time Green functions by the equation \cite{shabaev:2002:119}
\be \label{eq:2.2}
S^{\rm R}_{\gamma_f,A';\gamma_i,A} 
  &=& Z_3^{-1}\; \delta(k_f^0 - k_i^0)
  \oint_{\Gamma_A}dE'\oint_{\Gamma_A}dE\,g_{\gamma_f,A';\gamma_i,A}(E',E,k_i^0)\nonumber  \\[0.2cm]
  & & \hspace*{2.0cm} \times\; \Bigl[\frac{1}{2\pi i}\oint_{\Gamma_A}dE\,g_{A'A'}(E)\Bigr]^{-1/2}
  \Bigl[\frac{1}{2\pi i}\oint_{\Gamma_A}dE\,g_{AA}(E)\Bigr]^{-1/2}\,,
\ee
where the contour $\Gamma_{A}$ encloses the pole corresponding to the bound-electron states
with the energy $E_A$. This contour also excludes all further singularities of the Green functions
$g_{\gamma_f,A';\gamma_i,A}$, $g_{AA}$, and $g_{A'A'}$, which are defined in a similar way
as in Ref.~\cite{shabaev:2002:119}. The factor $Z_3$ is a renormalization constant for
the absorbed and emitted photons lines. The Green function $g_{\gamma_f,A';\gamma_i,A}(E',E,k_i^0)$
describes the scattering of a photon by bound electrons, while the Green functions
$g_{AA}(E)$ and $g_{A'A'}(E)$ characterize the initial and final bound-electron states.
Since the Rayleigh $S$-matrix element $S^{\rm R}_{\gamma_f,A';\gamma_i,A}$ is
expressed in terms of the Green functions, it can be calculated order-by-order by applying 
the QED perturbation theory with regard to the radiation-matter interaction.

In the following, we shall consider in further detail the non-resonant Rayleigh scattering of 
light by helium-like ions, i.e.\ for photon energies which are not close to possible 
excitations of any quasi-stationary bound state. The zeroth-order two-electron wave functions 
$u_A$ and $u_{A'}$ are constructed as linear combinations of Slater determinants, 
$A = (a_1,a_2)_{J_A M_A}$ and $A' = (a_1,a_2)_{J_{A'} M_{A'}}$ as
\be
u_A(\bfr_1,\bfr_2)
  = F_A \frac{1}{\sqrt{2}}\,
  \Biggl|\begin{array}{cc}
         u_{a_1}(\bfr_1) & u_{a_2}(\bfr_1) \\
         u_{a_1}(\bfr_2) & u_{a_2}(\bfr_2) \\
         \end{array}\Biggr|
 = F_A \frac{1}{\sqrt{2}}\sum_P (-1)^P |Pa_1 Pa_2 \ra\,,
\ee
where $F_A$ is a shorthand notation for the summation over the Clebsch-Gordan coefficients
\be
F_A |a_1 a_2 \ra = \sum_{m_{a_1},m_{a_2}} C^{J_A M_A}_{j_{a_1} m_{a_1} j_{a_2} m_{a_2}} |a_1 a_2 \ra
  \times\Biggl\{\begin{array}{cc}
  1\,,          & a_1\neq a_2 \\
  1/\sqrt{2}\,, & a_1  =  a_2 \\
                \end{array}\,,
\ee
$j_a$ the one-electron total angular momentum and $m_a$ its projection, $P$ is the
permutation operator, giving rise to the sign $(-1)^P$ of the permutation for any permutation
of the electron coordinates. The same notations hold for the final state $A'$. The one-electron 
wave functions $u_{a_1}$ and $u_{a_2}$ are found by solving the Dirac equation with the Coulomb 
potential of the nucleus.

\subsection{Zeroth-order approximation}
\label{sec:2A}

\begin{figure}
\includegraphics{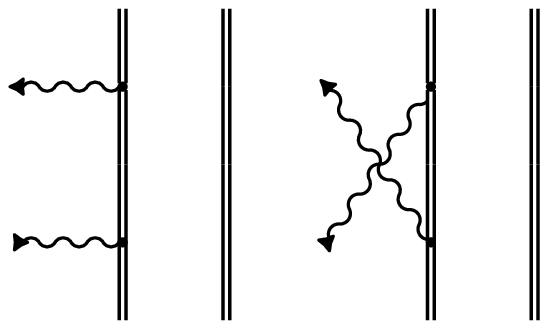}
\caption{Feynman diagrams of the Rayleigh scattering in zeroth-order approximation.
The double line indicates the electron propagators in the Coulomb field of the nucleus, 
while the photon absorption and emission are depicted by the wavy line with incoming and 
outgoing arrows, respectively.
\label{fig:1}}
\end{figure}

In order to calculate the $S$-matrix element $S^{\rm R}_{\gamma_f,A';\gamma_i,A}$
of the Rayleigh scattering according to Eq.~(\ref{eq:2.2}), we decompose the two-time
Green functions in a perturbation series with an expansion parameter $\alpha$ and
group the terms of the same order together. In zeroth-order approximation, the
corresponding Feynman diagrams are depicted in Fig.~\ref{fig:1}. Then, by employing
Eqs.~(\ref{eq:2.1}) and (\ref{eq:2.2}) we obtain the zeroth-order Rayleigh scattering
amplitude $\tau^{(0)}_{\gamma_f,A';\gamma_i,A}$ in a following form
\be \label{eq:2A.1}
\tau^{(0)}_{\gamma_f,A';\gamma_i,A} = \frac{1}{2\pi i}
  \oint_{\Gamma_A}dE' \oint_{\Gamma_A}dE\,
  g^{(0)}_{\gamma_f,A';\gamma_i,A}(E',E,k_i^0)\,,
\ee
where the superscript ``$(0)$'' indicates the order of the perturbation expansion.
According to the Feynman rules, the zeroth-order Green function $g^{(0)}_{\gamma_f,A';\gamma_i,A}$
can be written as
\be \label{eq:2A.2}
\lefteqn{ g^{(0)}_{\gamma_f,A';\gamma_i,A}(E',E,k_i^0)
  \delta(E'+k_f^0-k_i^0-E) }\nonumber\\
  &=&
  F_A F_{A'} \sum_{P,\,Q}(-1)^{P+Q} \int_{-\infty}^{\infty} dp_1^0 dp_2^0 dp_1'^0 dp_2'^0 dq^0
  \,\delta(E-p_1^0-p_2^0)\,\delta(E'-p_1'^0-p_2'^0)\nonumber\\
  &\times&
  \left\{\la Pa_1'|\frac{i}{2\pi}\sum_{n_1}\frac{|n_1\ra\la n_1|}{p_1'^0-u\veps_{n_1}}
  \frac{2\pi}{i}R^*_f\delta(p_1'^0+k_f^0-q^0)
  \frac{i}{2\pi}\sum_{n_2}\frac{|n_2\ra\la n_2|}{q^0-u\veps_{n_2}}
  \frac{2\pi}{i}R_i\delta(q^0-k_i^0-p_1^0)\right.\nonumber\\
  &\times&
  \frac{i}{2\pi}\sum_{n_3}\frac{|n_3\ra\la n_3|}{p_1^0-u\veps_{n_3}}|Qa_1\ra
  \la Pa_2'|\frac{i}{2\pi}\sum_{n_4}\frac{|n_4\ra\la n_4|}{p_2^0-u\veps_{n_4}}|Qa_2\ra
  \delta(p_2'^0-p_2^0)\nonumber\\
  &+&\la Pa_1'|\frac{i}{2\pi}\sum_{n_1}\frac{|n_1\ra\la n_1|}{p_1'^0-u\veps_{n_1}}
  \frac{2\pi}{i}R_i\delta(p_1'^0-k_i^0-q^0)
  \frac{i}{2\pi}\sum_{n_2}\frac{|n_2\ra\la n_2|}{q^0-u\veps_{n_2}}
  \frac{2\pi}{i}R^*_f\delta(q^0+k_f^0-p_1^0)\nonumber\\
  &\times&
  \left.\frac{i}{2\pi}\sum_{n_3}\frac{|n_3\ra\la n_3|}{p_1^0-u\veps_{n_3}}|Qa_1\ra
  \la Pa_2'|\frac{i}{2\pi}\sum_{n_4}\frac{|n_4\ra\la n_4|}{p_2^0-u\veps_{n_4}}|Qa_2\ra
  \delta(p_2'^0-p_2^0)\right\}\nonumber\\
  &=&
  \frac{i}{2\pi}\frac{\delta(E'+k_f^0-k_i^0-E)}{(E'-E_A^{(0)})(E-E_A^{(0)})}
  F_A F_{A'} \sum_{P,\,Q}(-1)^{P+Q} \sum_n
  \left\{\frac{\la Pa_1'|R^*_f|n\ra\la n|R_i|Qa_1\ra
         \delta_{Pa_2' Qa_2}}{E-\veps_{Qa_2}+k_i^0-u\veps_n}\right.\nonumber\\
  &+&
  \left.\frac{\la Pa_1'|R_i|n\ra\la n|R^*_f|Qa_1\ra
         \delta_{Pa_2' Qa_2}}{E'-\veps_{Pa_2}-k_i^0-u\veps_n}\right\}\,,
\ee
$R^*_f(\bfr) = e\,\balpha\,\beps_f^*\,e^{-i\bfk_f\bfr} / \sqrt{2k_f^0 (2\pi)^3}$
and $R_i(\bfr) = e\,\balpha\,\beps_i\,e^{i\bfk_i\bfr} / \sqrt{2k_i^0 (2\pi)^3}$ 
are the emission and absorption operators, $\balpha$ is the vector of the Dirac 
$\alpha$-matrices, and where the zeroth-order energy of the bound electrons 
$E_A^{(0)}$ is equal to the sum of the one-electron Dirac energies 
$E_A^{(0)}=\veps_{a_1}+\veps_{a_2}$. Furthermore, the factor $u = 1-i0$ 
in expression (\ref{eq:2A.2}) preserves the proper treatment of poles of the electron 
propagators. In the case of the non-resonant scattering, the expressions in the curly brackets 
of Eq.~(\ref{eq:2A.2}) are regular functions of $E$ and $E'$ as long as
$E \approx E_A^{(0)}$ and $E' \approx E_A^{(0)}$. By substituting Eq.~(\ref{eq:2A.2}) 
into Eq.~(\ref{eq:2A.1}) and by integrating over $E$ and $E'$, one thus
easily obtains
\be \label{eq:2A.3}
\tau^{(0)}_{\gamma_f,A';\gamma_i,A}
  &=& -F_A F_{A'} \sum_{P,\,Q}(-1)^{P+Q} \delta_{Pa_2' Qa_2} \sum_n
  \left\{\frac{\la Pa_1'|R^*_f|n\ra\la n|R_i|Qa_1\ra}{\veps_{Qa_1}+k_i^0-u\veps_n}
       + \frac{\la Pa_1'|R_i|n\ra\la n|R^*_f|Qa_1\ra}{\veps_{Pa_1}-k_i^0-u\veps_n}
  \right\}\nonumber\\
  &=& -F_A F_{A'} \sum_n
  \left\{\left(\frac{\la a_1'|R^*_f|n\ra\la n|R_i|a_1\ra}{\veps_{a_1}+k_i^0-u\veps_n}
             + \frac{\la a_1'|R_i|n\ra\la n|R^*_f|a_1\ra}{\veps_{a_1}-k_i^0-u\veps_n}
  \right)\delta_{m_{a_2'}m_{a_2} }\right.\nonumber\\
  &+&\left.\left(
              \frac{\la a_2'|R^*_f|n\ra\la n|R_i|a_2\ra}{\veps_{a_2}+k_i^0-u\veps_n}
            + \frac{\la a_2'|R_i|n\ra\la n|R^*_f|a_2\ra}{\veps_{a_2}-k_i^0-u\veps_n}
  \right)\delta_{m_{a_1'}m_{a_1} }\right\}\,.
\ee

Before we shall proceed further, let us discuss this expression for
the zeroth-order Rayleigh scattering amplitude $\tau^{(0)}_{\gamma_f,A';\gamma_i,A}$. 
Obviously, this amplitude splits into two pieces as indicated by the
round brackets. These pieces correspond to either the scattering by just the $a_1$
(first round brackets) or $a_2$ (second round brackets) electrons. In the zeroth-order
approximation, hence, the obtained Rayleigh scattering amplitude corresponds to 
the IPA formulas as they are widely used for the theoretical description of the 
Rayleigh scattering, see, e.g., Ref.~\cite{roy:1999:3}.

Using the (zeroth-order) Rayleigh scattering amplitude from above, the angle-differential 
Rayleigh scattering cross section defined by Eq.~(\ref{eq:2.0}) is given in zeroth-order 
approximation by
\be \label{eq:2A.4}
d\sigma^{(0)}(\bfk_f;\bfk_i,\beps_i) =
  \frac{(2\pi)^4 (k_i^0)^2}{2J_A+1}\sum_{M_A,\,M_{A'}} \sum_{\beps_f}\,
  |\tau^{(0)}_{\gamma_f,A';\gamma_i,A}|^2\,d\Omega_f\,,
\ee
while the corresponding zeroth-order total cross section gives rise to
\be \label{eq:2A.5}
\sigma^{(0)}(\bfk_i,\beps_i) =
  \frac{(2\pi)^4 (k_i^0)^2}{2J_A+1}\sum_{M_A,\,M_{A'}} \sum_{\beps_f} \int
  |\tau^{(0)}_{\gamma_f,A';\gamma_i,A}|^2\,d\Omega_f\,.
\ee

In the following, we shall go beyond the zeroth-order approximation and investigate 
the effects due to the electron-electron correlations. To do so, we need to account for 
the interelectronic-interaction correction to the Rayleigh scattering amplitude. In the next 
subsection we now present the corresponding formulas for the first-order 
interelectronic-interaction effects.

\begin{figure}
\includegraphics{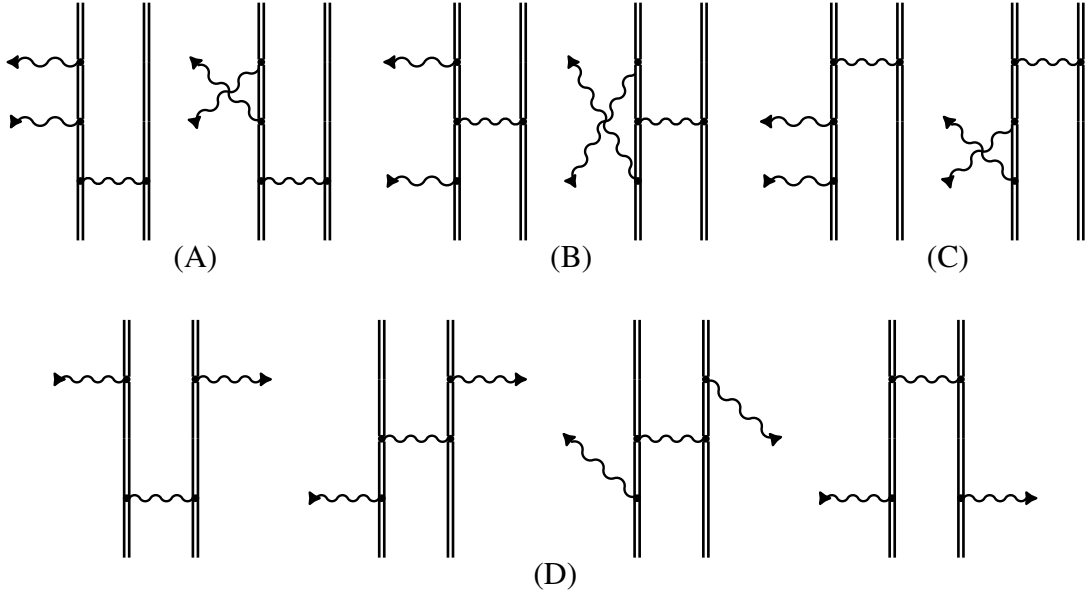}
\caption{Feynman diagrams to represent the first-order interelectronic-interaction
corrections to the Rayleigh scattering process. The internal wavy line stands for
the photon propagator. The rest notations are the same as in Fig.~\ref{fig:1}.
\label{fig:2}}
\end{figure}

\subsection{First-order interelectronic-interaction correction}
\label{sec:2B}

\begin{figure}
\includegraphics{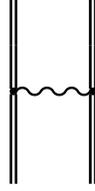}
\caption{One-photon exchange Feynman diagram. Notations are the same as in
Figs.~\ref{fig:1} and \ref{fig:2}.
\label{fig:3}}
\end{figure}

In order to obtain the expression for the first-order interelectronic-interaction
correction to the Rayleigh $S$-matrix element, we have to collect all first-order terms
in Eq.~(\ref{eq:2.2}) as they arise from the perturbation expansion of the Green
functions. Using Eq.~(\ref{eq:2.1}), the corresponding interelectronic-interaction
correction to the scattering amplitude $\Delta\tau^{(1)}_{\gamma_f,A';\gamma_i,A}$
is given by
\be \label{eq:2B.1}
\Delta\tau^{(1)}_{\gamma_f,A';\gamma_i,A} &=& \frac{1}{2\pi i}
  \left[\oint_{\Gamma_A}dE'\oint_{\Gamma_A}dE\,
  \Delta g^{(1)}_{\gamma_f,A';\gamma_i,A}(E',E,k_i^0)\right.\nonumber\\
  &-&\left.\frac{1}{2}\oint_{\Gamma_A}dE'\oint_{\Gamma_A}dE\,
  g^{(0)}_{\gamma_f,A';\gamma_i,A}(E',E,k_i^0)\;
  \left(\frac{1}{2\pi i}\oint_{\Gamma_A}dE\, \Delta g_{A'A'}^{(1)}(E)
  +\frac{1}{2\pi i}\oint_{\Gamma_A}dE\, \Delta g_{AA}^{(1)}(E)\right)\right]\,,
\ee
where the first-order interelectronic-interaction correction to the scattering
Green function $\Delta g^{(1)}_{\gamma_f,A';\gamma_i,A}$ can be represented
by Feynman diagrams as displayed in Fig.~\ref{fig:2}. The corresponding corrections
to the Green functions $\Delta g_{A'A'}^{(1)}$ and $\Delta g_{AA}^{(1)}$ are
defined by the first-order interelectronic-interaction diagram shown in
Fig.~\ref{fig:3}.
In addition, by making use of the Feynman rules for the two-time Green functions 
one can derive the final expressions for the scattering amplitude correction 
$\Delta\tau^{(1)}_{\gamma_f,A';\gamma_i,A}$. This derivation is technically
very similar to those as performed in Refs.~\cite{indelicato:2004:062506,
volotka:2011:062508}, where the interelectronic-interaction effects were investigated
for one- and two-photon bound-bound transitions in helium-like ions. For the sake
of brevity, we therefore omit here the cumbersome expressions and go on to the final
form of the first-order interelectronic-interaction correction 
$\Delta\tau^{(1)}_{\gamma_f,A';\gamma_i,A}$ which can be written as a following sum
\be \label{eq:2B.2}
\Delta\tau^{(1)}_{\gamma_f,A';\gamma_i,A} =
    \Delta\tau^{(1A)}_{\gamma_f,A';\gamma_i,A}
  + \Delta\tau^{(1B)}_{\gamma_f,A';\gamma_i,A}
  + \Delta\tau^{(1C)}_{\gamma_f,A';\gamma_i,A}
  + \Delta\tau^{(1D)}_{\gamma_f,A';\gamma_i,A}
  + \Delta\tau^{(1R)}_{\gamma_f,A';\gamma_i,A}\, .
\ee
Here, the different contributions are distinguished by the superscripts $(1A)$, $(1B)$, 
$(1C)$ and $(1D)$ and correspond to the so-called \textit{irreducible} parts of the diagrams
as shown in Fig.~\ref{fig:2} (A), (B), (C), and (D), respectively. These terms are given by
the expressions:
\be \label{eq:2B.3}
\Delta\tau^{(1A)}_{\gamma_f,A';\gamma_i,A} &=& -F_A F_{A'} \sum_{P,\,Q}(-1)^{P+Q}
  \sum_{n_1,n_2}^{\veps_{n_2} \ne \veps_{Pa_1}}\left\{
  \frac{\la Pa_1'|R^*_f|n_1 \ra \la n_1|R_i|n_2 \ra \la n_2 Pa_2'|I(\veps_{Pa_2}-\veps_{Qa_2})|Qa_1 Qa_2 \ra}
       {(\veps_{Pa_1}+k_i^0-u\veps_{n_1})(\veps_{Pa_1}-\veps_{n_2})}\right.\nonumber\\
  &+&\left.
  \frac{\la Pa_1'|R_i|n_1 \ra \la n_1|R^*_f|n_2 \ra \la n_2 Pa_2'|I(\veps_{Pa_2}-\veps_{Qa_2})|Qa_1 Qa_2 \ra}
       {(\veps_{Pa_1}-k_i^0-u\veps_{n_1})(\veps_{Pa_1}-\veps_{n_2})}\right\}\,.
\ee
\be \label{eq:2B.4}
\Delta\tau^{(1B)}_{\gamma_f,A';\gamma_i,A} &=& -F_A F_{A'} \sum_{P,\,Q}(-1)^{P+Q}
  \sum_{n_1,n_2}\left\{
  \frac{\la Pa_1'|R^*_f|n_1 \ra \la n_1 Pa_2'|I(\veps_{Pa_2}-\veps_{Qa_2})|n_2 Qa_2 \ra \la n_2|R_i|Qa_1 \ra}
       {(\veps_{Pa_1}+k_i^0-u\veps_{n_1})(\veps_{Qa_1}+k_i^0-u\veps_{n_2})}\right.\nonumber\\
  &+&\left.
  \frac{\la Pa_1'|R_i|n_1 \ra \la n_1 Pa_2'|I(\veps_{Pa_2}-\veps_{Qa_2})|n_2 Qa_2 \ra \la n_2|R^*_f|Qa_1 \ra}
       {(\veps_{Pa_1}-k_i^0-u\veps_{n_1})(\veps_{Qa_1}-k_i^0-u\veps_{n_2})}\right\}\,,
\ee
\be \label{eq:2B.5}
\Delta\tau^{(1C)}_{\gamma_f,A';\gamma_i,A} &=& -F_A F_{A'} \sum_{P,\,Q}(-1)^{P+Q}
  \sum_{n_1,n_2}^{\veps_{n_1} \ne \veps_{Qa_1}}\left\{
  \frac{\la Pa_1' Pa_2'|I(\veps_{Pa_2}-\veps_{Qa_2})|n_1 Qa_2 \ra \la n_1|R^*_f|n_2 \ra \la n_2|R_i|Qa_1 \ra}
       {(\veps_{Qa_1}-\veps_{n_1})(\veps_{Qa_1}+k_i^0-u\veps_{n_2})}\right.\nonumber\\
  &+&\left.
  \frac{\la Pa_1' Pa_2'|I(\veps_{Pa_2}-\veps_{Qa_2})|n_1 Qa_2 \ra \la n_1|R_i|n_2 \ra \la n_2|R^*_f|Qa_1 \ra}
       {(\veps_{Qa_1}-\veps_{n_1})(\veps_{Qa_1}-k_i^0-u\veps_{n_2})}\right\}\,,
\ee
\be \label{eq:2B.6}
\Delta\tau^{(1D)}_{\gamma_f,A';\gamma_i,A} &=& -F_A F_{A'} \sum_{P,\,Q}(-1)^{P+Q}
  \sum_{n_1,n_2}\left\{
  \frac{\la Pa_1'|R_i|n_1 \ra \la Pa_2'|R^*_f|n_2 \ra \la n_1 n_2|I(k_i^0-\veps_{Pa_1}+\veps_{Qa_1})|Qa_1 Qa_2 \ra}
       {(\veps_{Pa_1}-k_i^0-u\veps_{n_1})(\veps_{Pa_2}+k_i^0-u\veps_{n_2})}\right.\nonumber\\
  &+&
  \frac{\la Pa_1' n_2|I(k_i^0-\veps_{Pa_1}+\veps_{Qa_1})|n_1 Qa_2 \ra \la Pa_2'|R^*_f|n_2 \ra \la n_1|R_i|Qa_1 \ra}
       {(\veps_{Qa_1}+k_i^0-u\veps_{n_1})(\veps_{Pa_2}+k_i^0-u\veps_{n_2})}\nonumber\\
  &+&
  \frac{\la Pa_1' n_2|I(k_i^0+\veps_{Pa_1}-\veps_{Qa_1})|n_1 Qa_2 \ra \la Pa_2'|R_i|n_2 \ra \la n_1|R^*_f|Qa_1 \ra}
       {(\veps_{Qa_1}-k_i^0-u\veps_{n_1})(\veps_{Pa_2}-k_i^0-u\veps_{n_2})}\nonumber\\
  &+&\left.
  \frac{\la Pa_1' Pa_2'|I(k_i^0-\veps_{Pa_1}+\veps_{Qa_1})|n_1 n_2 \ra \la n_1|R_i|Qa_1 \ra \la n_2|R^*_f|Qa_2 \ra}
       {(\veps_{Qa_1}+k_i^0-u\veps_{n_1})(\veps_{Qa_2}-k_i^0-u\veps_{n_2})}\right\}\,,
\ee
and where $I(\omega) = e^2\alpha^{\mu}\alpha^{\nu}D_{\mu \nu}(\omega)$ refers to the interelectronic-interaction
operator with the photon propagator $D_{\mu \nu}(\omega)$. The diagrams displayed in Fig.~\ref{fig:2}(A) and (C)
contain also the so-called \textit{reducible} parts, while this is not the case for the diagrams shown in
Fig.~\ref{fig:2}(B) and (D). Thus, the last term in Eq.~(\ref{eq:2B.2}) $\Delta\tau^{(1R)}_{\gamma_f,A';\gamma_i,A}$
is the total reducible contribution, which arises from the second term in the square brackets of Eq.~(\ref{eq:2B.1})
and from the reducible parts of the diagrams (A) and (C) in Fig.~\ref{fig:2}. The total reducible term can be
written as
\be \label{eq:2B.7}
\Delta\tau^{(1R)}_{\gamma_f,A';\gamma_i,A}
  &=& F_A F_{A'} \sum_{P,\,Q}(-1)^{P+Q} \delta_{Pa_2' Qa_2} \sum_n
  \left\{\left(\frac{\la Pa_1'|R^*_f|n\ra\la n|R_i|Qa_1\ra}{(\veps_{Qa_1}+k_i^0-u\veps_n)^2}
             + \frac{\la Pa_1'|R_i|n\ra\la n|R^*_f|Qa_1\ra}{(\veps_{Pa_1}-k_i^0-u\veps_n)^2}\right)
         \Delta E_A^{(1)}\right.\nonumber\\
  &+&\left.\left(\frac{\la Pa_1'|R^*_f|n\ra\la n|R_i|Qa_1\ra}{\veps_{Qa_1}+k_i^0-u\veps_n}
             + \frac{\la Pa_1'|R_i|n\ra\la n|R^*_f|Qa_1\ra}{\veps_{Pa_1}-k_i^0-u\veps_n}\right)
  \Delta E_A^{\pr(1)}\right\}\,,
\ee
where
\be
\Delta E_A^{(1)} = F_A F_{\tilde{A}}\sum_{P}(-1)^P
  \la P\tilde{a}_1 P\tilde{a}_2|
    I(\veps_{P\tilde{a}_1}-\veps_{a_1})
  |a_1 a_2 \ra
\ee
is the one-photon exchange correction and
\be
\Delta E_A^{\pr(1)} = F_A F_{\tilde{A}}\sum_{P}(-1)^P
  \la P\tilde{a}_1 P\tilde{a}_2|
    \frac{dI(x)}{dx}\Big|_{\Delta=\veps_{P\tilde{a}_1}-\veps_{a_1}}
  |a_1 a_2 \ra
\ee
is the first-order derivative of the one-photon exchange correction.

Finally, the angle-differential Rayleigh scattering cross section up to the 
first-order in the interelectronic interaction is given by
\be \label{eq:2B.8}
d\sigma^{(1)}(\bfk_f;\bfk_i,\beps_i) =
  \frac{(2\pi)^4 (k_i^0)^2}{2J_A+1}\sum_{M_A,\,M_{A'}} \sum_{\beps_f}\,
  |\tau^{(0)}_{\gamma_f,A';\gamma_i,A} + \Delta\tau^{(1)}_{\gamma_f,A';\gamma_i,A}|^2\,d\Omega_f\,,
\ee
while the corresponding total cross section takes the form
\be \label{eq:2B.9}
\sigma^{(1)}(\bfk_i,\beps_i) =
  \frac{(2\pi)^4 (k_i^0)^2}{2J_A+1}\sum_{M_A,\,M_{A'}} \sum_{\beps_f} \int
  |\tau^{(0)}_{\gamma_f,A';\gamma_i,A} + \Delta\tau^{(1)}_{\gamma_f,A';\gamma_i,A}|^2\,d\Omega_f\,.
\ee

These Rayleigh scattering cross sections take rigorously into account the many-electron 
effects on a level of the one-photon exchange. These calculations enable us to analyze 
systematically the importance of the many-electron effects upon the elastic scattering 
cross sections. 
However, before we continue with this analysis let us consider a different 
approach for including the many-electron effects in the next subsection.

\subsection{Screening potential approximation}
\label{sec:2C}

Apart from the rigorous approach above, the zeroth-order or independent-particle approximation
of Eq.~(\ref{eq:2A.3}) also facilitates a partial account of the interelectronic-interaction
effects in the Rayleigh cross sections. To this end, we shall start from an extended
Furry picture in which a central screening potential is incorporated into
the zeroth-order Hamiltonian. In this case, the formula (\ref{eq:2A.3}) for the zeroth-order 
Rayleigh amplitude remains formally the same, while the initial, intermediate, and final
one-electron wave functions are now generated within a \textit{mean-field} potential
by including, in addition to the Coulomb field, also some screening potential. 
In practice, however, this approach only includes some (major) parts of the
many-electron effects for highly charged ions. Here, we shall separate this part 
from the complete first-order result in Eq.~(\ref{eq:2B.2}) by restricting
ourselves to the static Coulomb part in the photon propagator $D_{\mu\nu}$ as well as
to the spherical terms in its multipole expansion. Below, we shall refer to this approximation
by the superscript "scr" in the Rayleigh scattering amplitude
\be \label{eq:2C.1}
\Delta\tau^{(1){\rm scr}}_{\gamma_f,A';\gamma_i,A} =
    \Delta\tau^{(1A){\rm scr}}_{\gamma_f,A';\gamma_i,A}
  + \Delta\tau^{(1B){\rm scr}}_{\gamma_f,A';\gamma_i,A}
  + \Delta\tau^{(1C){\rm scr}}_{\gamma_f,A';\gamma_i,A}
  + \Delta\tau^{(1R){\rm scr}}_{\gamma_f,A';\gamma_i,A}\,.
\ee
Let us mention here that the contribution of the diagrams (D) in Fig.~\ref{fig:2} 
should also be excluded in this case since these diagrams are of inherent many-electron
character. In the leading order this approximation is equivalent to the IPA with
the Dirac-Hartree-Fock potential. The difference arisen from the higher-order terms
can be neglected for highly charged ions.

With these "screening" corrections to the Rayleigh scattering amplitude, the 
angle-differential and total cross sections take now the form
\be \label{eq:2C.2}
d\sigma^{\rm scr}(\bfk_f;\bfk_i,\beps_i) =
  \frac{(2\pi)^4 (k_i^0)^2}{2J_A+1}\sum_{M_A,\,M_{A'}} \sum_{\beps_f}\,
  |\tau^{(0)}_{\gamma_f,A';\gamma_i,A} + \Delta\tau^{(1){\rm scr}}_{\gamma_f,A';\gamma_i,A}|^2\,d\Omega_f\,,
\ee
and
\be \label{eq:2C.3}
\sigma^{\rm scr}(\bfk_i,\beps_i) =
  \frac{(2\pi)^4 (k_i^0)^2}{2J_A+1}\sum_{M_A,\,M_{A'}} \sum_{\beps_f} \int
  |\tau^{(0)}_{\gamma_f,A';\gamma_i,A} + \Delta\tau^{(1){\rm scr}}_{\gamma_f,A';\gamma_i,A}|^2\,d\Omega_f\,,
\ee
respectively. Since this approximation can be obtained from Eq.~(\ref{eq:2A.3}) by just 
making use of a screening potential in solving the Dirac equation, we shall refer to
it as the IPA. --- In the next section, we now discuss the numerical procedure and the 
methods employed in the computation of the cross sections.

\section{Computations}
\label{sec:3}

The formulas (\ref{eq:2A.3}), (\ref{eq:2B.2}), and (\ref{eq:2C.1}) for the transition amplitudes,
as obtained in the previous section, require further simplifications to make
detailed computations feasible. For instance, in order to perform the angular integrations 
in the one- and two-electron matrix elements, we utilize the well-known multipole expansion 
technique. Indeed, this angular integration can be carried out analytically by expanding 
the transition operators $R_f^*$ and $R_i$ as well as the photon propagator $D_{\mu\nu}$ in 
multipole series. For the sake of brevity, we do not recall the corresponding expressions here 
and just refer the reader to the literature for further details \cite{varshalovich}. 
The infinite multipole summations over the incoming and outgoing photon multipoles are 
further restricted by analyzing the convergence and more often than not we summed
up to 10 multipoles.

Numerically most demanding in the computations is the infinite summation over the
complete Dirac spectrum $n_1$ and $n_2$, which not only contain the bound states 
but also the positive- and negative-energy Dirac continuum. In order to perform such
a summation several independent approaches were employed previously in the 
consideration of the Rayleigh scattering. One method formulated in 
Ref.~\cite{brown:1954:51} is based on a solution of an inhomogeneous Dirac equation, a
method that was found quite successful and was utilized in a good number of
calculations of the elastic scattering cross sections \cite{brenner:1954:59,brown:1956:387,
brown:1957:89,johnson:1976:692,kissel:1980:1970,kane:1986:75,roy:1986:1178}.
Another approach is known as the finite basis-set method. This technique enables one
to replace an infinite summation in the spectral representation of the electron
propagator by a summation over a finite basis set and was utilized for
calculating the Rayleigh scattering in Refs.~\cite{safari:2012:043405,safari:2015:271}. 
Still another approach is based on the exact Dirac-Coulomb Green's function
which can be represented in a closed form as a superposition of the regular and 
irregular solutions of the Dirac equation. In the case of a point-like Coulomb potential,
for example, it can be expressed analytically \cite{wichmann:1956:843}.
In Refs.~\cite{surzhykov:2013:062515,surzhykov:2015:144015} this method was also utilized
in the calculation of the Rayleigh scattering cross sections.

\begin{figure}
\includegraphics[scale=0.59]{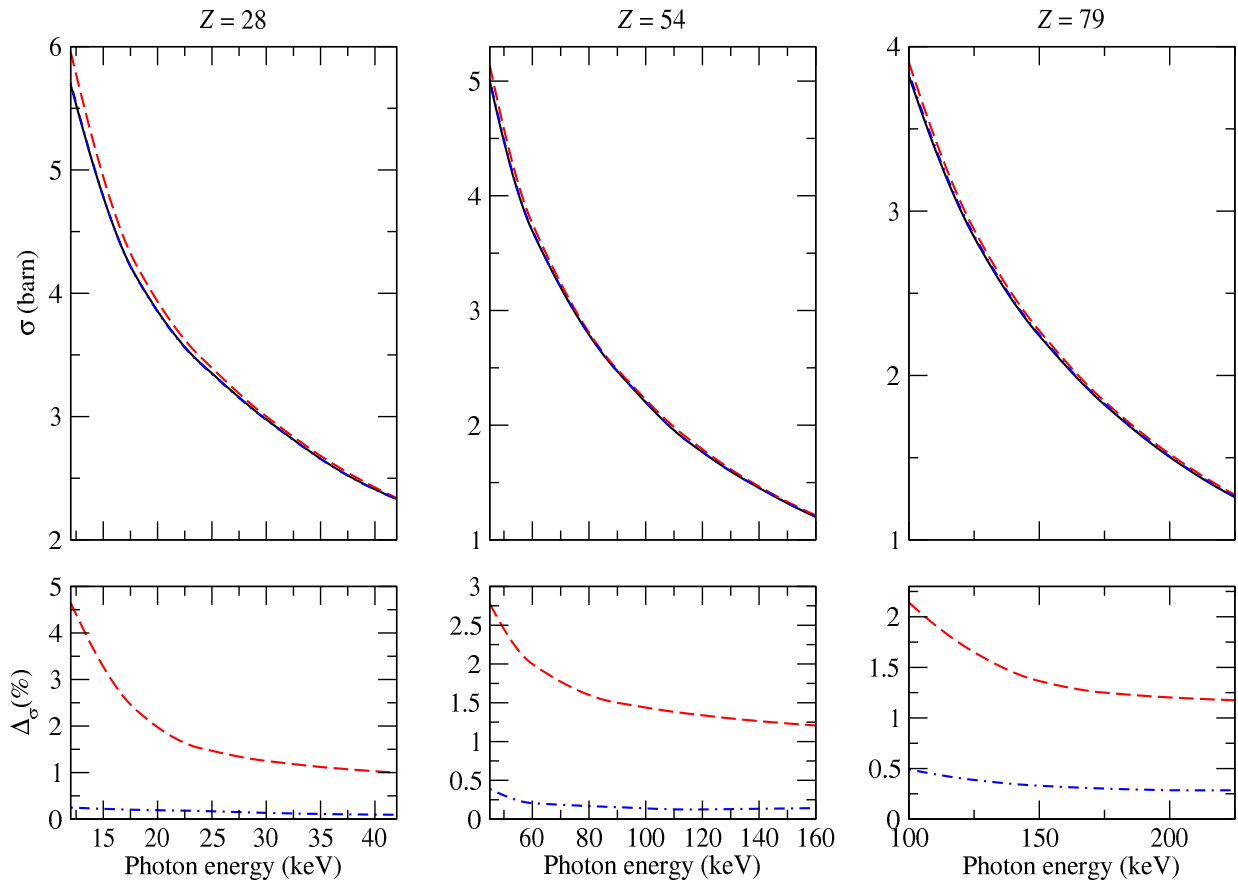}
\caption{(Color online) Total cross section for the Rayleigh scattering of x rays
by helium-like Ni$^{26+}$ ($Z=28$), Xe$^{52+}$ ($Z=54$) and Au$^{77+}$ ($Z=79$) ions in
their $^1S_0$ ground state and as function of the photon energy. Predictions are 
compared for three different approximations in the evaluation of the scattering amplitudes: 
A pure Coulomb potential $\sigma^{(0)}$ (red dashed line), 
use of a screening potential $\sigma^{\rm scr}$ (blue dash-dotted line)
as well as by applying the full many-electron procedure in $\sigma^{(1)}$ (black solid line). 
The lower panel displays the differences (in \%{}) between the Coulomb
$\Delta_\sigma^{(0)}$ (red dashed line) and the screening cross sections $\Delta_\sigma^{\rm scr}$
(blue dash-dotted line), each relative to the many-electron computations.
\label{fig:tot}}
\end{figure}

In the present work, we made use of the two latter approaches: the finite basis-set 
method and the Dirac-Coulomb Green's function. The finite basis set was constructed from
$B$-splines \cite{sapirstein:1996:5213} by employing the dual-kinetic-balance 
approach \cite{shabaev:2004:130405}. In addition, the analytic Dirac-Coulomb 
Green's function for a point-like Coulomb potential was employed in terms of 
the Whittaker functions \cite{yerokhin:1999:800}. The finite basis set method 
enables us to considerably reduce the numerical effort in the calculations. 
The reason for this is the separation of the radial variables and the subsequent 
integrations of the single radial integrals. However, when the energies of 
the incident photons are larger than the ionization threshold, the application
of the finite basis-set technique is critically hampered in some of the terms 
(diagrams). For example, this is the case for the first term of Eq.~(\ref{eq:2A.3})
with the denominator $(\veps_{Qa_1}+k_i^0-u\veps_n)$. This term has a pole 
in the energy continuum (spectrum) at $\veps_n = \veps_{Qa_1}+k_i^0+i0$
that cannot be treated accurately in any finite basis-set method due to the
discretized spectrum and the summation over just a finite number of basis functions.
In contrast, the use of the Dirac-Coulomb Green's function is free of such
difficulties as it represents the exact electron propagator. Therefore, all
the electron propagators with energies $\veps_{a_1}+k_i^0$ or $\veps_{a_2}+k_i^0$ 
were treated by means of the Dirac-Coulomb Green's function $G_{\kappa_n}$:
\be
\sum_n \frac{|n \ra \la n|}{\veps_a+k_i^0-u\veps_n} \equiv
  \sum_{n_r,\,\kappa_n} \frac{u_{n_r\kappa_n}(\bfr_1) u^\dagger_{n_r\kappa_n}(\bfr_2)}{\veps_a+k_i^0-u\veps_{n_r\kappa_n}} =
  \sum_{\kappa_n} G_{\kappa_n}(\veps_a+k_i^0,\bfr_1,\bfr_2)\,,
\ee
and where $n_r$ and $\kappa_n$ denote the principal and Dirac angular quantum numbers,
respectively. For all other propagators, the finite basis set representation were
employed. This combination of different (numerical) techniques enables us to substantially
reduce the computational time, while we still obtain (very) accurate results.

Still, a quite serious problem occurs for those terms where \textit{two}
Dirac-Coulomb Green's functions appear in the calculations. For example, such a
term is the first contribution in Eq.~(\ref{eq:2B.4}) which can be re-written 
by means of the Dirac-Coulomb Green's functions as
\be
\tau^{(1B)}_{\gamma_f,A';\gamma_i,A} &=& -F_A F_{A'} \sum_{P,\,Q}(-1)^{P+Q} \sum_{\kappa_{n_1},\,\kappa_{n_2}}
  \int d\bfr_1 d\bfr_2 d\bfr_3 d\bfr_4\,
   u^\dagger_{Pa_1'}(\bfr_1) u^\dagger_{Pa_2'}(\bfr_4)
   R^*_f(\bfr_1)
   G_{\kappa_{n_1}}(\veps_{Pa_1}+k_i^0,\bfr_1,\bfr_2)\nonumber\\
   &\times&
   I(\veps_{Pa_2}-\veps_{Qa_2},\bfr_2,\bfr_4)
   G_{\kappa_{n_2}}(\veps_{Qa_1}+k_i^0,\bfr_2,\bfr_3)
   R_i(\bfr_3)
   u_{Qa_1}(\bfr_3)u_{Qa_1}(\bfr_4) + {\rm second\; term}\,.
\ee
Here, the integration over the $\bfr_2$ coordinate involves the two Dirac-Coulomb
Green's functions $G_{\kappa_{n_1}}(\veps_{Pa_1}+k_i^0,\bfr_1,\bfr_2)$ and
$G_{\kappa_{n_2}}(\veps_{Qa_1}+k_i^0,\bfr_2,\bfr_3)$, and together with 
the photon propagator $I(\veps_{Pa_2}-\veps_{Qa_2},\bfr_2,\bfr_4)$. In contrast to
the integrations over the coordinates $\bfr_1$, $\bfr_3$, and $\bfr_4$, no
bound-electron wave function is involved here. Therefore, the integral
over $r_2$ converges only very slowly for large values of $r_2$, and 
this makes a straightforward numerical integration extremely cumbersome.
For this reason, we here applied the method of the complex-plane rotation of
the coordinate integration contour and which was previously used in 
studying bremsstrahlung \cite{yerokhin:2010:062702} and double photoionization 
processes \cite{yerokhin:2011:032703}. We have introduced the radius of the atom $R$,
i.e.\ an radius outside of which all the bound-state electron wave functions vanish.
In the integrals over $r_1$, $r_3$, and $r_4$ we then set the upper bound to $R$ and
split the remaining over $r_2$ into two domains $[0,R]$ and $[R, R-i\infty)$. With
these rearrangements in the integration procedure, the integral over the second domain
decays exponentially and does no longer cause any (numerical) problem.

In order to check the consistency of our numerical results we have performed
calculations in different gauges, namely in length and velocity gauge
for the photon wave functions and in Coulomb and Feynman gauge of the
photon propagator. The obtained results were found in perfect agreement,
independently of the chosen gauge form.

\begin{figure}
\includegraphics[scale=0.9]{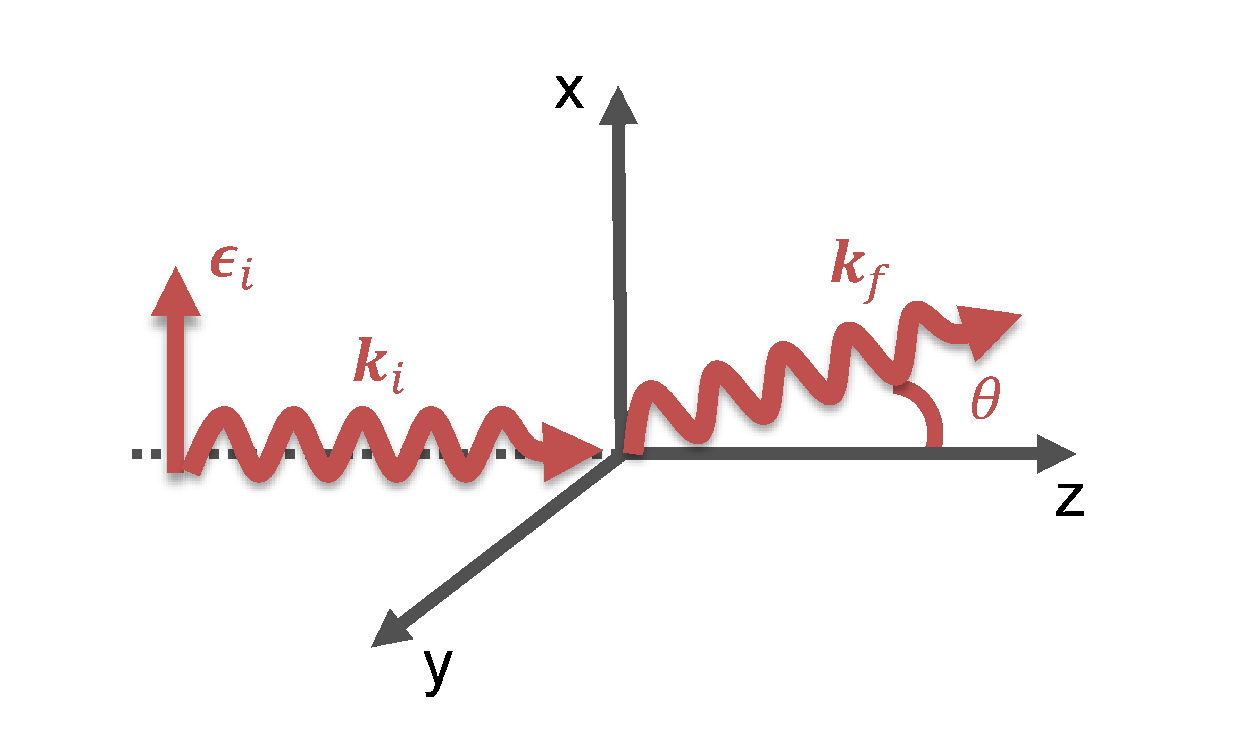}
\caption{(Color online) Geometry for describing the Rayleigh scattering 
of photons at a nucleus which is taken as the origin of the coordinates. 
The $z$-axis is chosen along the direction of the incoming light, 
while the reaction plane is defined by the $xz$ plane. 
The scattering angle $\theta$ then uniquely defines the direction 
of the scattered photon within the reaction plane.
\label{fig:geometry}}
\end{figure}

\begin{figure}
\includegraphics[scale=0.59]{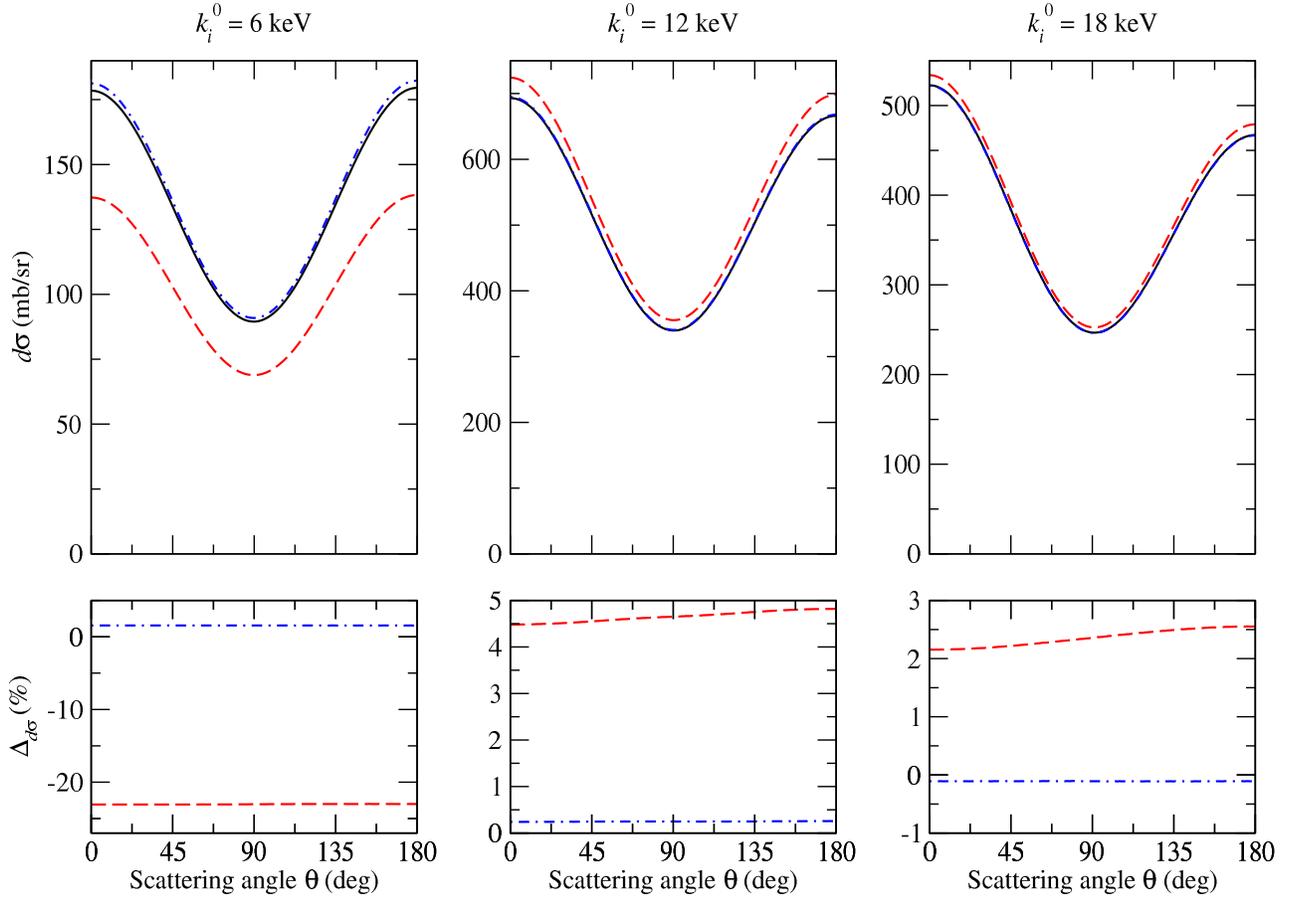}
\caption{(Color online) Angle-differential cross section for the Rayleigh scattering 
of unpolarized incoming photons with energies $k^0_i = 6$ keV (left column), 
$k^0_i = 12$ keV (middle column) and $k^0_i = 18$ keV (right column) by helium-like
Ni$^{26+}$ ions in their ground state. Theoretical results are shown for the three
approximations in this work: Coulomb $d\sigma^{(0)}$ (red dashed line), screening 
$d\sigma^{\rm scr}$ (blue dash-dotted line) as well as the many-electron computations
$d\sigma^{(1)}$ (black solid line). In the lower panel, again the differences (in \%{}) 
between the Coulomb $\Delta_{d\sigma}^{(0)}$ (red dashed line) and screening 
$\Delta_{d\sigma}^{\rm scr}$ results (blue dash-dotted line) are shown, just relative 
to the many-electron computations.
\label{fig:28_unpol}}
\end{figure}

\begin{figure}
\includegraphics[scale=0.59]{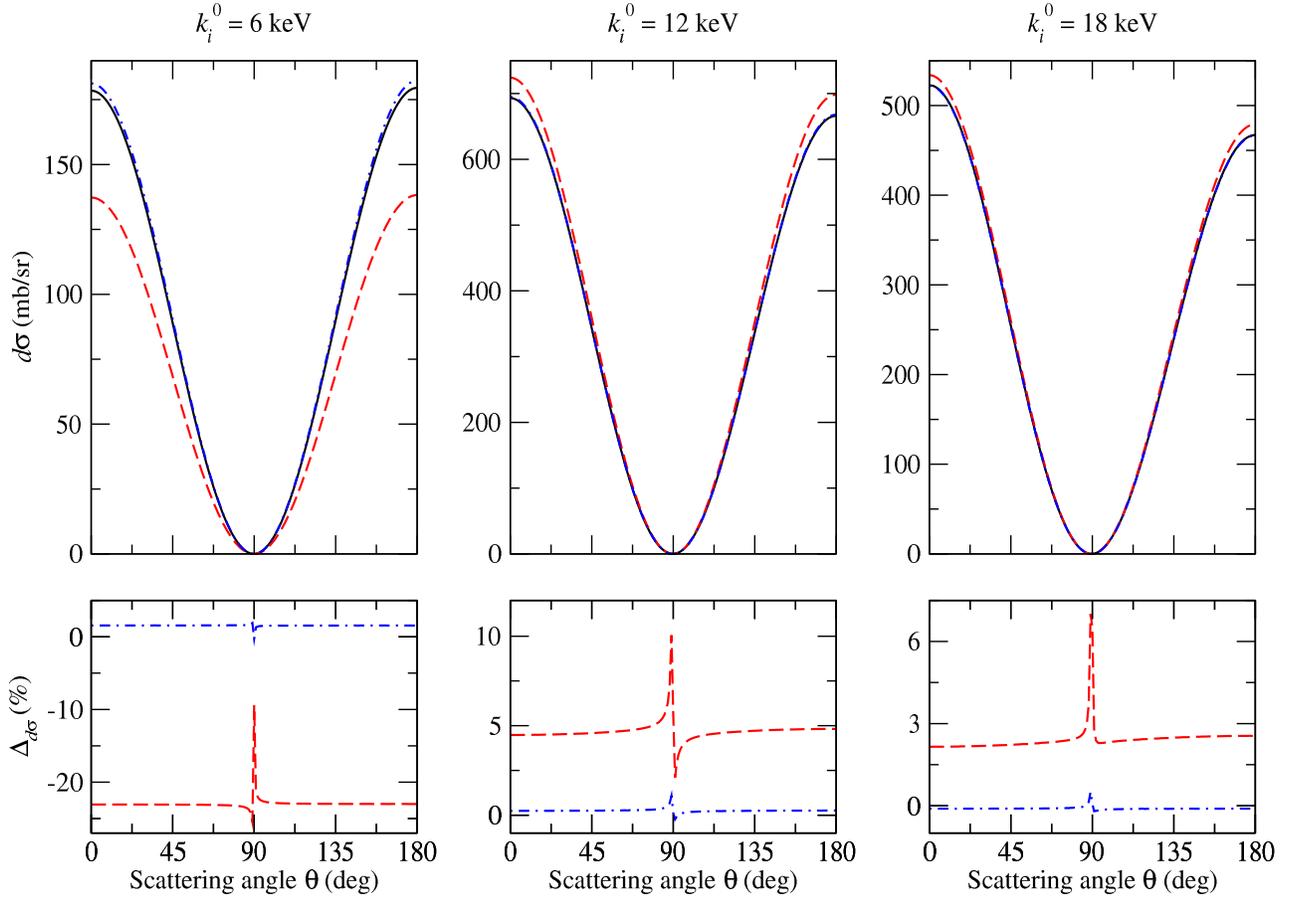}
\caption{(Color online) The same as Fig.~\ref{fig:28_unpol} but for linearly-polarized
incoming photons.
\label{fig:28_pol}}
\end{figure}

\section{Results and discussion}
\label{sec:4}

Although the formulas in Sec.~\ref{sec:2} applies generally for helium-like
ions, independent of their particular state, detailed calculations have been performed 
here only for the Rayleigh scattering by such ions in their $1s^2\;\,^1S_0$ 
ground state. In particular, we here aim to investigate the "many-electron"
effects beyond the IPA, for which two set of calculations were carried out: A first one,
in which the zeroth-order approximation with the Coulomb wave functions were applied in
Eqs.~(\ref{eq:2A.4}) and (\ref{eq:2A.5}) and to which we refer below as the Coulomb results.
A second IPA computation was carried out by using the screening potential approximation and 
Eqs.~(\ref{eq:2C.2}) and (\ref{eq:2C.3}); these computations are referred to as the 
screening results. Both of these IPA calculations are compared with the complete zeroth- 
and first-orders results as obtained by means of Eqs.~(\ref{eq:2B.8}) and 
(\ref{eq:2B.9}), called the many-electron treatment below.

Fig.~\ref{fig:tot} displays  the total cross section for the Rayleigh scattering 
of light by helium-like Ni$^{26+}$ ($Z=28$), Xe$^{52+}$ ($Z=54$) and Au$^{77+}$ ($Z=79$) 
ions and as function of the photon energy. Results are shown for photon energies well 
above the ionization threshold and for the three computations above: 
Coulomb $\sigma^{(0)}$, screening $\sigma^{\rm scr}$, and many-electron $\sigma^{(1)}$.
To make the differences in the theoretical predictions more explicit, the relative deviations
(in \%{}) with regard to the many-electron data are also shown in the lower panel of this
figure: $\Delta_\sigma^{(0)} = (\sigma^{(0)}-\sigma^{(1)}) / \sigma^{(1)}$ and
$\Delta_\sigma^{\rm scr} = (\sigma^{\rm scr}-\sigma^{(1)}) / \sigma^{(1)}$. As seen from
this figure, all three computations give quite similar results with slightly larger
deviations near to the ionization threshold. At the threshold, the Rayleigh scattering
cross sections appear to be very sensitive due to differences in the calculated threshold
energies in the three approximations. For the scattering on helium-like Au$^{77+}$ ions,
for example, the calculated threshold energies are 93.411 keV, 91.857 keV and 91.656 keV
for the Coulomb, screening, and the many-electron computations, respectively. While, in
the Coulomb approximation, the ionization threshold is completely determined by the
one-electron Dirac binding energy, the interelectronic interactions modify this value
in the screening and many-electron calculations. The difference between the ionization
energies in the screening and many-electron approximation is due to the contribution
of the Breit interaction to the ground-state energy. Thus, the behaviour of the
differences $\Delta_\sigma^{(0)}$ and $\Delta_\sigma^{\rm scr}$ near to the ionization
threshold as a function of the nuclear charge can be understood from the ratio of the
one-photon exchange and Breit interaction contribution to the threshold energy, respectively.
The first ratio decreases while the second one increases with increasing nuclear charge.
For large photon energies, all approximations slowly converge each other. This behaviour
was expected at high energies \cite{roy:1999:3} since the binding effects become less and
less important with increasing of the energy of scattered photon. The similar finding was
also obtained for the case of the helium atom in Ref.~\cite{lin:1975:1946}.

In addition to the total cross sections, we shall now investigate the angular distribution
of the scattered photons and for which we need to fix a geometry for describing the scattering
process [cf.~Fig.~\ref{fig:geometry}]. In the Furry picture, as mentioned above, we can treat
the photon scattering in the rest-frame of the nucleus, taken as the origin of the coordinates. 
Moreover, we can choose the $z$-axis along the wave vector $\bfk_i$ of the incident radiation.
Than the scattered photon wave vector $\bfk_f$ is completely described by the two angles:
azimuthal $\theta$ and polar $\phi$. What concerns the polarization of the incident photon,
we consider here two scenarios. In the first scenario the incoming light is completely
unpolarized, and the angular distribution is independent on the polar angle $\phi$. In the
second scenario the incoming light is completely linearly polarized along the $x$-axis, than
one observes a dependence of the scattering cross section on the polar angle $\phi$ as well.
Here, we restrict ourselves to the geometry when the photon is observed at the angle $\phi=0$,
thus, defining the reaction plane to be the $xz$ plane.

\begin{figure}
\includegraphics[scale=0.59]{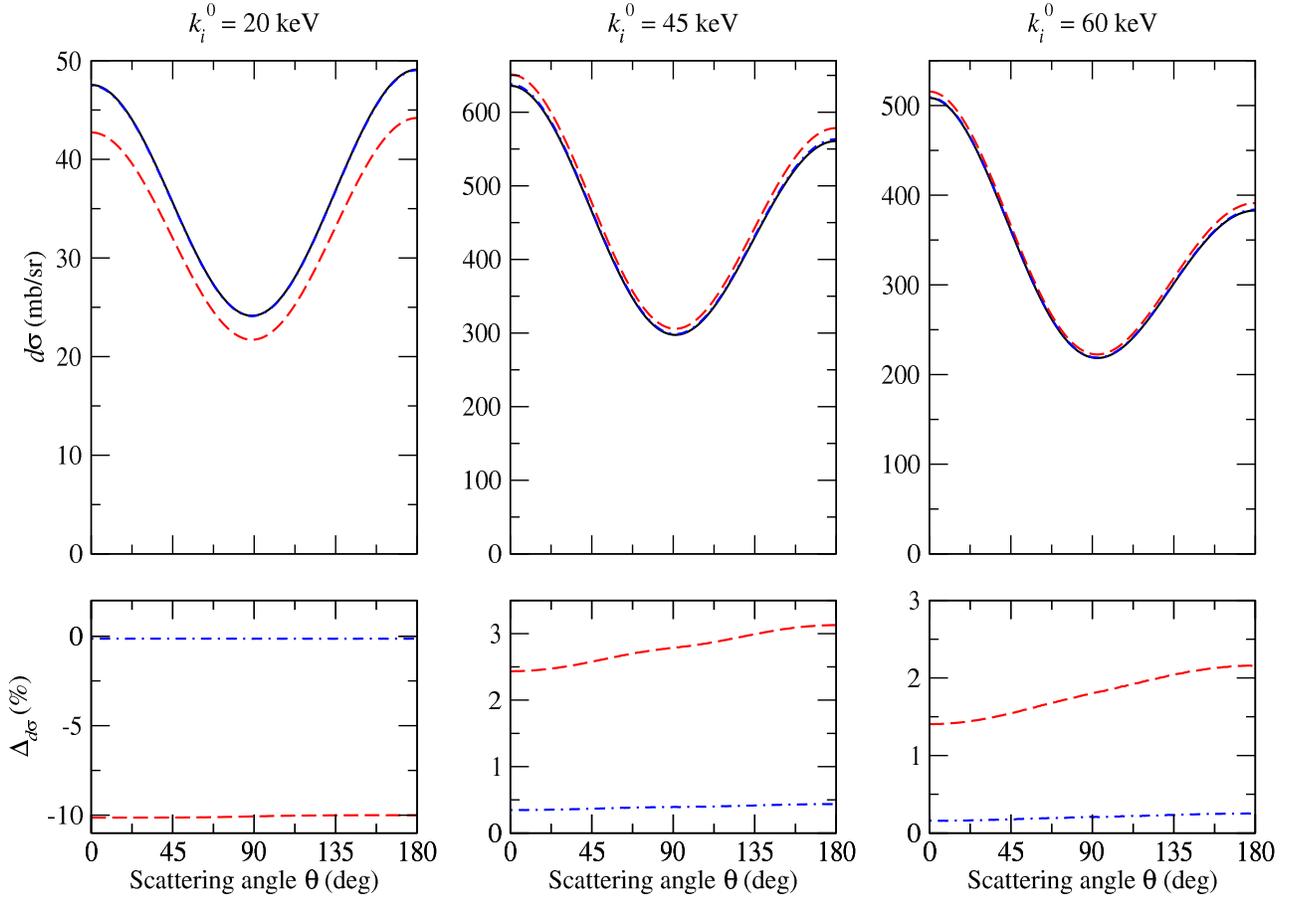}
\caption{(Color online) The same as Fig.~\ref{fig:28_unpol} but for the Rayleigh scattering
by helium-like Xe$^{52+}$ ions and for incident photon energies $k^0_i = 20$ keV (left column), 
$k^0_i = 45$ keV (middle column) and $k^0_i = 60$ keV (right column), respectively.
\label{fig:54_unpol}}
\end{figure}

\begin{figure}
\includegraphics[scale=0.59]{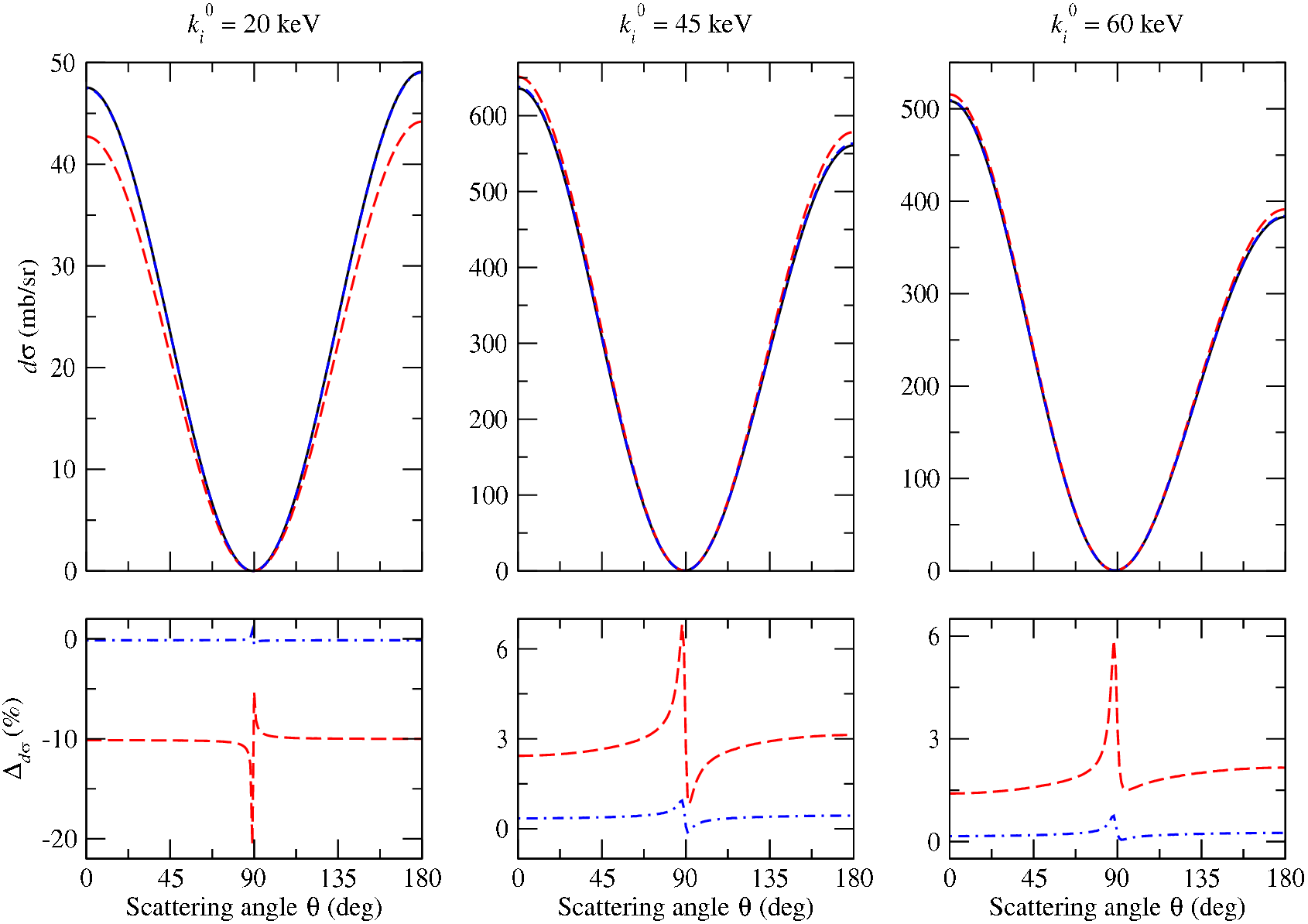}
\caption{(Color online) The same as Fig.~\ref{fig:54_unpol} but for linearly-polarized incoming photons.
\label{fig:54_pol}}
\end{figure}

\begin{figure}
\includegraphics[scale=0.59]{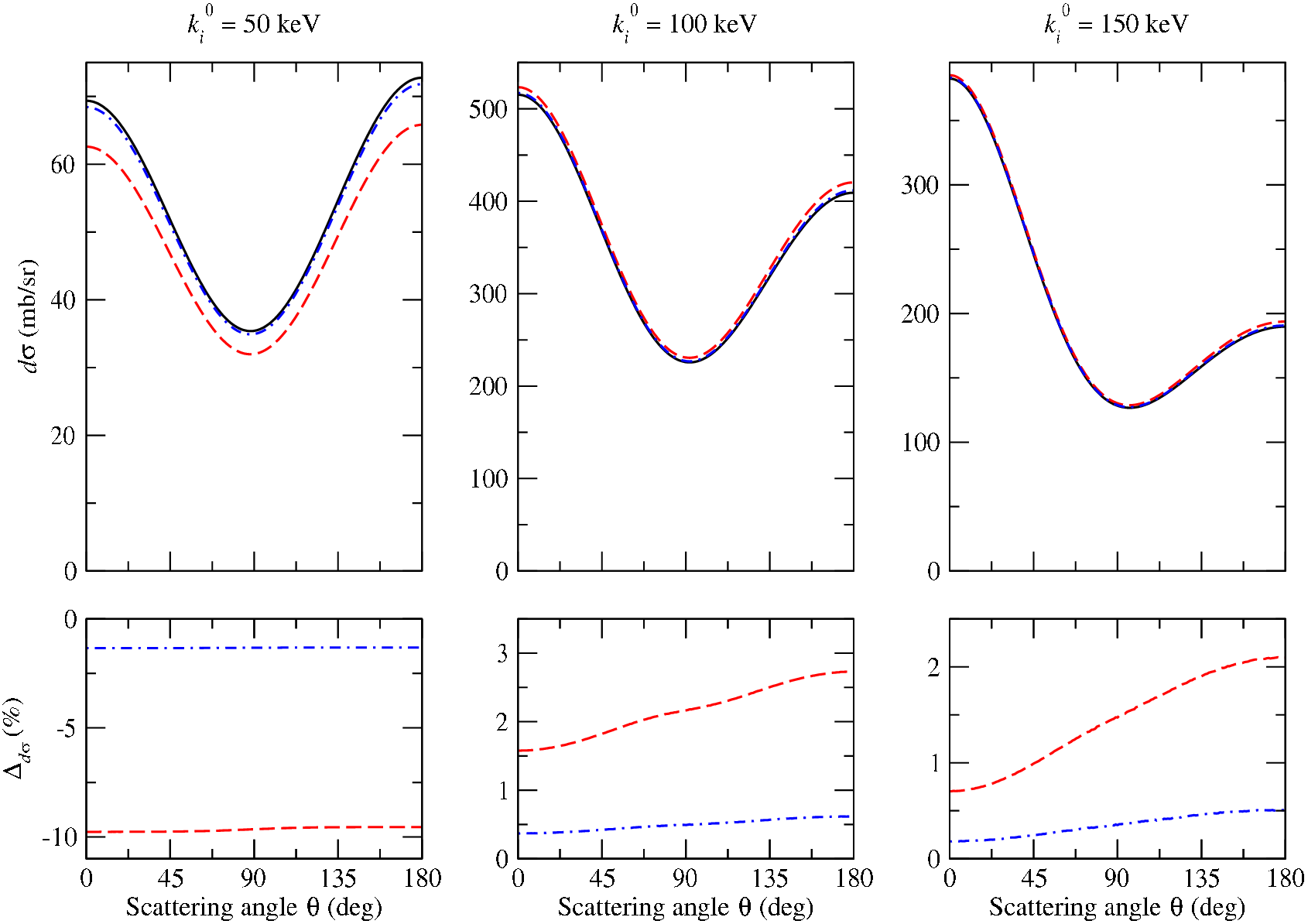}
\caption{(Color online) The same as Fig.~\ref{fig:28_unpol} but for the Rayleigh scattering
by helium-like Au$^{77+}$ ions and for incident photon energies $k^0_i = 50$ keV (left column), 
$k^0_i = 100$ keV (middle column) and $k^0_i = 150$ keV (right column), respectively.
\label{fig:79_unpol}}
\end{figure}

\begin{figure}
\includegraphics[scale=0.59]{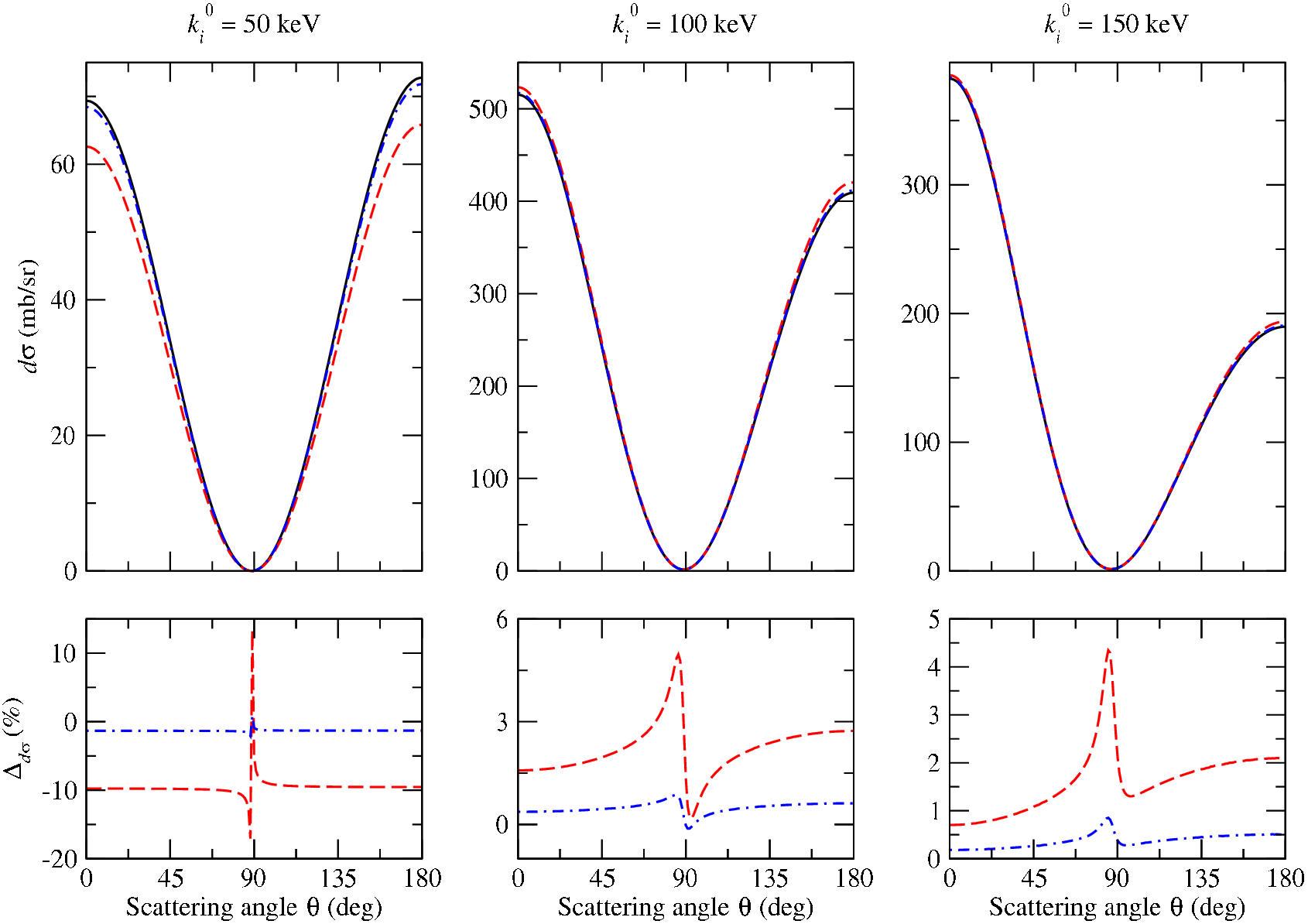}
\caption{(Color online) The same as Fig.~\ref{fig:79_unpol} but for linearly-polarized incoming photons.
\label{fig:79_pol}}
\end{figure}

In Figs.~\ref{fig:28_unpol} and \ref{fig:28_pol}, we display the angle-differential 
cross sections for the Rayleigh scattering of unpolarized and linearly polarized 
incoming light, if scattered by helium-like Ni$^{26+}$ ions in their 
ground state. Analogue computations were performed also for helium-like Xe$^{26+}$ 
and Au$^{77+}$ ions and are given in Figs.~\ref{fig:54_unpol}-\ref{fig:79_pol}, 
respectively. These angular distributions of the Rayleigh-scattered photons are
shown for three energies of the incoming photons, namely for about half of the 
ionization threshold as well as for photon energies that are 10\% and 50\% larger that
this threshold. A different behaviour of the angular distributions for 
a different polarization of the incident light can be expected already
from the nonrelativistic electric-dipole approximation which predicts a
$1+cos^2\theta$ shape for unpolarized (incoming) light and a $cos^2\theta$ shape
for completely polarized radiation, respectively. For large photon energies, 
however, \textit{non-dipole} effects become also important and lead to (more or 
less) strong deviations from this electric-dipole behaviour above \cite{kissel_pratt}.

Similar as for the total Rayleigh cross sections, we have analyzed and compare 
in Figs.~\ref{fig:28_unpol}-\ref{fig:79_pol} the angle-differential cross sections
in the three approximations above: Coulomb $d\sigma^{(0)}$, screening $d\sigma^{\rm scr}$,
and many-electron $d\sigma^{(1)}$. Since the deviations between the different computations
are typically small over a wide range of angle, we also display the relative differences,
$\Delta_{d\sigma}^{(0)} = (d\sigma^{(0)}-d\sigma^{(1)}) / d\sigma^{(1)}$ and 
$\Delta_{d\sigma}^{\rm scr} = (d\sigma^{\rm scr}-d\sigma^{(1)}) / d\sigma^{(1)}$,
i.e.\ normalized to the many-electron data as our best approximation. These
relative differences are shown in the lower panels of these figures. While these
differences are typically small for energies well above the ionization threshold
(in the right columns of these figures), sizeable deviations arise below of this 
threshold, and especially for the Coulomb results. The relative differences are
large at low $Z$ and clearly demonstrate the importance of the 
interelectronic-interaction effects. The deviations are however less 
significant between the screening and many-electron computations, i.e.\ less than
2\% for all the cases considered here. This demonstrates that the
interelectronic-interaction effects can be treated quite efficiently within the
IPA with a screening potential for the case of the Rayleigh scattering of photons
on helium-like ions in their ground state.

\section{Summary and outlook}
\label{sec:5}

In summary, a systematic QED treatment has been presented for the first-order corrections
of the interelectronic interaction to the Rayleigh scattering of photons by helium-like ions.
By applying and comparing three different theoretical and computational approaches, we rigorously
explore the role of the many-electron effects beyond the IPA to the Rayleigh scattering by
highly charged ions. Detailed calculations for the total and angle-differential cross sections
were evaluated for the scattering by helium-like Ni$^{26+}$, Xe$^{52+}$, and Au$^{77+}$ ions
in their $1s^2\;\,^1S_0$ ground state and, especially, for unpolarized and completely-polarized
incident radiation. We here found that the interelectronic-interaction effects are more important
for photon energies below and just above of the ionization threshold. However, the major part of
these many-electron contributions can be taken into account by IPA calculations with a screening
potential.

For photon energies well above the valence and sub-valence binding energies, this conclusion can
be generalized also towards more complex atoms. For such atoms and ions, the Rayleigh scattering
is typically dominated by the scattering from the inner-shell electrons. However, for photon
energies compared with the valence and sub-valence binding energies, the Rayleigh scattering
cross sections are strongly affected by the scattering from the outer-shell electrons and 
further care has to be taken in employing the IPA. This can be proven for forward emissions
($\theta = 0$), where the Rayleigh scattering cross section can be related to the photoionization
cross section via the dispersion relation and the optical theorem \cite{zhou:1992:2983}.
In complex atoms the photoionization cross section near to the ionization threshold often shows
various structures, such as the Cooper minima and resonances due to inner-shell excitations,
which appear to be very sensitive to many-electron effects. These electronic-structure phenomena
will likely affect also the Rayleigh scattering cross sections, especially, for the cases of
the scattering by the open-shell electrons \cite{zhou:1992:6906}. The theoretical formalism above
can be applied to such open-shell atoms, though (much) further effort will be needed for its
efficient numerical implementation.

\section*{Acknowledgments}

This work has been supported by the BMBF (Grant No. 05K13VHA) and RFBR (Grant No. 16-02-00538).
V.A.Y. acknowledges support by the Ministry of Education and Science of Russian Federation
(program for organizing and carrying out scientific investigations).


\end{document}